\documentclass[12pt]{article}
\usepackage{hyperref}
\usepackage{amsmath,amssymb}
\usepackage{graphicx}
\usepackage{color}
\definecolor{myblue}{rgb}{0.14,0.11,0.49}
\definecolor{myred}{rgb}{0.74,0.22,0.15}
\definecolor{mygreen}{rgb}{0.05,0.52,0.42}
\definecolor{myyellow}{rgb}{0.96,0.92,0.13}
\definecolor{myorange}{rgb}{1,0.61,0.36}
\definecolor{mypurple}{rgb}{0.71,0.02,1}
\definecolor{noir}{gray}{0.} 

\newcommand{\Couleur}[1]{\textcolor{noir}{#1}}

\definecolor{htc}{rgb}{1,1,1} 

\def\be{\begin{equation}}
\def\ee{\end{equation}}
\def\bea{\begin{eqnarray}}
\def\eea{\end{eqnarray}}
\def\bc{\begin{center}}
\def\ec{\end{center}}
\def\bi{\begin{itemize}}
\def\ei{\end{itemize}}
\def\bs{\begin{slide}}
\def\es{\end{slide}}
\def\dd{\mathrm{d}}
\def\iC{\mathrm{i}}
\def\noi{\noindent}
\title{An explicit representation for the axisymmetric solutions of the free Maxwell equations}
\author{
Mayeul Arminjon\\
\small\it Univ. Grenoble Alpes, CNRS, Grenoble INP
, 3SR, F-38000 Grenoble, France\\
\small\it  E-mail: Mayeul.Arminjon@3sr-grenoble.fr
} 
\date{}
\begin{document}

\maketitle

Short title: {\it Explicit representation for axisymmetric free Maxwell fields}

\begin{abstract}

\noi Garay-Avenda\~no \& Zamboni-Rached (2014) defined two classes of axisymmetric solutions of the free Maxwell equations. We prove that the linear combinations of these two classes of solutions cover all totally propagating time-harmonic axisymmetric free Maxwell fields --- and hence, by summation on frequencies, all totally propagating axisymmetric free Maxwell fields. It provides an explicit representation for these fields. This will be important, e.g., to have the interstellar radiation field in a disc galaxy modelled as an exact solution of the free Maxwell equations.\\

\noi {\bf Keywords:} Maxwell equations; axial symmetry; exact solutions; electromagnetic duality.\\


\end{abstract}
\section{Introduction}

Axially symmetric solutions of the Maxwell equations are quite important, at least as an often relevant approximation. For instance, axisymmetric magnetic fields occur naturally as produced by systems possessing an axis of revolution, such as disks or coils \cite{Garrett1951, Boridy1989}, or astrophysical systems like accretion disks \cite{Lovelace-et-al1990} or disk galaxies \cite{BeckWielebinsky2013}. Axisymmetric solutions are also used to model EM beams and their propagation (e.g. \cite{Nesterov-Niziev2001, Borghi-et-al2002}). In particular, non-diffracting beams are usually endowed with axial symmetry, see e.g. Refs. \cite{Durnin1987, Durnin_et_al1987, ZR_et_al2008}. Naturally, one often considers time-harmonic solutions, since a general time dependence is got by summing such solutions. Two classes of time-harmonic axisymmetric solutions of the free Maxwell equations, mutually associated by EM duality, have been introduced recently \cite{GAZR2014}. The main aim was to ``describe in exact and analytic form the propagation of nonparaxial scalar and electromagnetic beams." However, as noted by the authors of Ref. \cite{GAZR2014}, the analytical expression for a totally propagating time-harmonic axisymmetric solution $\Psi $ of the scalar wave equation, from which they start [Eq. (\ref{psi_monochrom}) below], covers all such solutions \cite{ZR_et_al2008} --- thus not merely ones corresponding to nonparaxial scalar beams. The first class of EM fields defined in Ref. \cite{GAZR2014} is obtained by associating with any such scalar solution $\Psi $ a vector potential ${\bf A}$ by Eq. (\ref{A_from_Az}) below, and the second class is deduced from the first one by EM duality \cite{GAZR2014}.  \\

The aim of the present work is to show that, by combining these two classes, one is able to describe actually {\it all} totally propagating time-harmonic axisymmetric EM fields --- and thus, by summing on frequencies, all totally propagating axisymmetric EM fields. To do that, we shall prove the following \hyperref[Theorem]{Theorem}: Any time-harmonic axisymmetric EM field (whether totally propagating or not) is the sum of two EM fields, say $\ ({\bf E}_1, {\bf B}_1)\ $ and $\ ({\bf E}'_2, {\bf B}'_2)\,$, deduced from two time-harmonic axisymmetric solutions of the scalar wave equation, say $\ \Psi _1\ $ and $\ \Psi _2\,$. The first EM field derives from the vector potential $\ {\bf A}_1 = \Psi _1 {\bf e}_z\ $, and the second one is deduced by EM duality from the EM field that derives from the vector potential $\ {\bf A}_2 = \Psi _2 {\bf e}_z$. Because this result is based on determining two vector potentials from merely two scalar fields, it was not necessarily expected. \\

The paper is organized as follows. Section \ref{scal_to_Max} presents and comments the results of Ref. \cite{GAZR2014}, and somewhat extends them, in particular by noting that Eqs. \ref{Bfi})--(\ref{Ez}) apply to any time-harmonic axisymmetric solution of the scalar wave equation. Also, Eqs. (\ref{Bphi_mono})--(\ref{Ez_poly}) are new. Section \ref{Max_to_scal} contains the main results of this work --- it gives the proof of the \hyperref[Theorem]{Theorem}. That proof is not immediate but uses standard mathematics with which one is familiar from classical field theory. In Section \ref{Conclusion}, we summarize our main results and we note that they lead straightforwardly to a method to get an explicit representation for all totally propagating axisymmetric solutions of the free Maxwell equations.

\section{From scalar waves to Maxwell fields} \label{scal_to_Max}
  \subsection{Axially symmetric scalar waves}

We adopt cylindrical coordinates $\rho ,\phi ,z$ about the symmetry axis, that is the $z$ axis. Any totally propagating, time-harmonic, axisymmetric solution of the scalar wave equation (of d'Alembert) can be written \cite{ZR_et_al2008, GAZR2014} as a sum of scalar Bessel beams:
\be\label{psi_monochrom}
\Couleur{\Psi _{\omega\ S} \,(t,\rho,z) = e^{-\iC \omega t} \int _{-K} ^{+K}\ J_0\left(\rho \sqrt{K^2-k^2}\right )\ e^{\iC k \, z} \,S(k)\, \dd k},
\ee
with $\omega $ the angular frequency, $K:=\omega /c$,
\footnote{
Beware that instead $K:=2\omega /c$ in Ref. \cite{GAZR2014}. Our notation seems as natural and gives more condensed formulas.
} 
and $J_0$ the first-kind Bessel function of order $0$. (Here $c$ is the velocity of light.) The Bessel beams were first introduced by Durnin \cite{Durnin1987}. A physical discussion of these beams and their ``non-diffracting" property can be found in Ref. \cite{McGloinDholakia2005}. On the other hand, a ``totally propagating" solution of the wave equation is one that does not have any evanescent mode. Loosely speaking, an evanescent mode can be described as a wave that behaves as a plane wave in some spatial direction, though with an imaginary wavenumber so that its amplitude decreases exponentially. See e.g. Refs. \cite{Jackson1998a, MikkiAntar2015}. In the present case, the totally propagating character of the wave (\ref{psi_monochrom}) means precisely that the axial wavenumber $k=k_z$ verifies $-K \le k \le K$, \cite{GAZR2014}, so that the radial wavenumber $k_\rho = \sqrt{K^2-k^2}$ is real as is also $k_z$. \\

Thus the (axial) ``wave vector spectrum" \Couleur{$S$ is a (generally complex) function of the real variable $k=k_z \quad (-K \leq k\leq  K )$}, that is the projection of the wave vector on the \Couleur{$z$} axis. This function $S$ determines the spatial dependency of the time-harmonic solution (\ref{psi_monochrom}) in the two-dimensional space left by the axial symmetry, i.e. the half-plane ($\rho \ge 0$, $z \in ]-\infty, +\infty [$). Thus, any totally propagating, time-harmonic, axisymmetric scalar wave $\Psi $ can be put in the explicit form (\ref{psi_monochrom}), in which no restriction has to be put on the ``wave vector spectrum" $S$ (except for a minimal regularity ensuring that the function $\Psi $ is at least twice continuously differentiable: the integrability of $S$, $S\in \mathrm{L}^1([-K, +K])$, would be enough for this). Of course, the {\it general} totally propagating axisymmetric solution of the scalar wave equation can be got from (\ref{psi_monochrom}) by an appropriate summation over a frequency spectrum: an integral (inverse Fourier transform) in the general case, or a discrete sum if a discrete frequency spectrum $(\omega _j)\ (j=1,...,N_\omega )$ is considered for simplicity:
\be\label{psi_avec_spectre}
\Couleur{\Psi (t,\rho,z) = \sum _{j=1} ^{N_\omega } \,\Psi _{\omega_j\ S_j}\,(t,\rho,z)},
\ee
where, for $j=1,...,N_\omega $, $\Psi _{\omega_j\ S_j}$ is the time-harmonic solution (\ref{psi_monochrom}), corresponding with frequency $\omega _j$ and wave vector spectrum $S_j$. The different weights $w_j$ which may be affected to the different frequencies can be included in the functions $S_j$, replacing $S_j $ by $ w_j S_j$.
 


  \subsection{Reminder on time-harmonic free Maxwell fields}

In this subsection, we recall equations more briefly recalled in Ref. \cite{GAZR2014}. 
The electric and magnetic fields in SI units are given in terms of the scalar and (3-)vector potentials $V$ and ${\bf A}$ by
\be\label{E_from_V&A}
{\bf E}= - \nabla V - \frac{\partial {\bf A}}{\partial t},
\ee
\be\label{B_from_A}
{\bf B}=  \mathrm{rot} {\bf A}.
\ee
These equations imply that ${\bf E}$ and ${\bf B}$ obey the first group of Maxwell equations. If one imposes the Lorenz gauge condition
\be\label{Lorenz}
\frac{1}{c^2}\frac{\partial V}{\partial t} + \mathrm{div} {\bf A} =0,
\ee
then the validity of the second group of the Maxwell equations in free space for ${\bf E}$ and ${\bf B}$ is equivalent to ask that  $V$ and ${\bf A}$ verify d'Alembert's wave equation \cite{L&L, Jackson1998b}. Moreover, if one assumes a harmonic time-dependence for $V$ and ${\bf A}$:
\be\label{V&A_harmonic}
V(t,{\bf x}) = e^{-\iC \omega t}\,\hat{V}({\bf x}),\qquad  {\bf A}(t,{\bf x}) = e^{-\iC \omega t}\,\hat{{\bf A}}({\bf x}),
\ee
then the wave equation for ${\bf A}$ becomes the Helmholtz equation:
\footnote{\ 
Of course, just the same equation (\ref{Helmholtz}) applies to the ``amplitude field" $\hat{{\bf A}}$.
}
\be\label{Helmholtz}
\Delta {\bf A} + \frac{\omega ^2}{c^2} {\bf A} = {\bf 0},
\ee 
and the Lorenz gauge condition (\ref{Lorenz}) rewrites as
\be\label{Lorenz_harmonic}
V = -\iC \,\frac{c^2}{\omega } \mathrm{div} {\bf A}.
\ee
If ${\bf A}$ is time-harmonic [Eq. (\ref{V&A_harmonic})$_2$] and obeys (\ref{Helmholtz}), then $V$ given by (\ref{Lorenz_harmonic}) is time-harmonic and  automatically satisfies the wave equation. The electric field (\ref{E_from_V&A}) is then easily rewritten as
\be\label{E_from_A}
{\bf E} = \iC \, \omega {\bf A} + \iC\,\frac{c^2}{\omega } \nabla \left ( \mathrm{div} {\bf A} \right ).
\ee
Thus, the data of a time-harmonic vector potential ${\bf A}$ obeying the wave equation, or equivalently obeying Eq. (\ref{Helmholtz}), determines a unique solution of the free Maxwell equations, by Eqs. (\ref{B_from_A}) and (\ref{E_from_A}), and that solution is time-harmonic with the same frequency $\omega $ as for ${\bf A}$.

\subsection{Time-harmonic axisymmetric fields: from a scalar wave to a Maxwell field}

To any solution $\Psi (t,\rho ,z)=e^{-\iC \omega t} \hat{\Psi } (\rho ,z)$ of the scalar wave equation having the form (\ref{psi_monochrom}), the authors of Ref. \cite{GAZR2014} associate a vector potential ${\bf A}$ by 
\be\label{A_from_Az}
\Couleur{{\bf A} := \Psi {\bf e}_z,\qquad \mathrm{or}\quad A_z :=\Psi , \ A_\rho  = A_\phi  =0}. 
\ee
(We shall denote by $({\bf e}_\rho , {\bf e}_\phi , {\bf e}_z)$ the standard point-dependent orthonormal basis associated with the cylindrical coordinates.) The form (\ref{psi_monochrom}) applies, as we mentioned, to totally propagating, axisymmetric, time-harmonic solutions of the scalar wave equation. Thus, in the way recalled in the foregoing subsection, they define a unique solution of the free Maxwell equations, which is time-harmonic. Their equations for the different components of this solution $({\bf E}, {\bf B})$ are as follows:
\bea\label{Bfi}
\Couleur{B_\phi} & = & \Couleur{-\frac{\partial A_z}{\partial \rho }},\qquad \Couleur{E_\phi =0},\\
\nonumber\\
\label{Erho}
\Couleur{E_\rho} & = & \Couleur{\iC \frac{c^2}{\omega } \frac{\partial^2 A_z}{\partial \rho \,\partial z}},\qquad \Couleur{B_\rho =0},\\
\nonumber\\
\label{Ez}
\Couleur{E_z} & = & \Couleur{\iC \frac{c^2}{\omega } \frac{\partial^2 A_z}{\partial z^2 } + \iC \omega A_z}, \qquad \Couleur{B_z=0}.
\eea
These equations follow easily from Eqs. (\ref{B_from_A}), (\ref{E_from_A}) and (\ref{A_from_Az}),  and from the axisymmetry of $A_z=\Psi (t,\rho ,z)$, by using the standard formulas for the curl and divergence in cylindrical coordinates. Equations (\ref{Bfi})--(\ref{Ez}) provide an axisymmetric EM field whose electric field is radially polarized (${\bf E} = E_\rho {\bf e}_\rho + E_z {\bf e}_z$), in short a ``radially polarized" EM field. \\

An ``azimuthally polarized" solution $({\bf E}', {\bf B}')$ (in the sense that ${\bf E}' = E'_\phi  {\bf e}_\phi $) of the free Maxwell equations can alternatively be deduced from the data $\Psi $, by transforming the solution (\ref{Bfi})--(\ref{Ez}) through the EM duality, that is:
\be\label{dual}
{\bf E}'=c{\bf B}, \quad {\bf B}' = -{\bf E}/c.
\ee

\vspace{2mm}
Now we observe this: the fact that the function $\Psi $ have the form (\ref{psi_monochrom}) plays no role in the derivation of the exact solution (\ref{Bfi})--(\ref{Ez}) to the free Maxwell equations. The only relevant fact is that $\Psi =\Psi(t,\rho ,z)$ is a time-harmonic axisymmetric solution of the scalar wave equation. Thus, with any axisymmetric time-harmonic solution $\Psi $ of the scalar wave equation, we can associate \hypertarget{GAZR}{two axisymmetric time-harmonic solutions} of the free Maxwell equations: the solution (\ref{Bfi})--(\ref{Ez}) and the one deduced from it by the duality (\ref{dual}). These two solutions will be called here the ``GAZR1 solution" and the ``GAZR2 solution", respectively, because both were derived in Ref. \cite{GAZR2014} --- although this was then for a totally propagating solution having the form (\ref{psi_monochrom}), and we have just noted that this is not necessary. \\

As usual, it is implicit that, in Eqs. (\ref{Bfi})--(\ref{Ez}),  $B_\phi$, $E_\rho$ and $E_z$ are actually the {\it real parts} of the respective r.h.s. [as are also ${\bf E}$ and ${\bf B}$ in Eqs. (\ref{E_from_V&A}), (\ref{B_from_A}), and (\ref{E_from_A})]. Thus, if $A_z$ is totally propagating and hence may be written in the form (\ref{psi_monochrom}), we obtain using the fact that $\dd J_0 /\dd x = -J_1(x)$:
\be\label{Bphi_mono}
B_{\phi \,\omega \, S} = {\mathcal Re} \left[e^{-\iC  \omega t} \int_{-K} ^{+K} \sqrt{K^2-k^2}\, J_1\left(\rho \sqrt{K^2-k^2}\right ) \,S(k) \,e^{ikz} \dd k \right ],
\ee
\be\label{Erho_mono}
E_{\rho \, \omega \, S} = {\mathcal Re} \left[-\iC \frac{c^2}{\omega } e^{-\iC  \omega t} \int_{-K} ^{+K} \sqrt{K^2-k^2}\, J_1\left(\rho \sqrt{K^2-k^2}\right )\iC k \,S(k) \,e^{ikz} \dd k \right ],
\ee
\be\label{Ez_mono}
E_{z \, \omega \, S} = {\mathcal Re} \left[\iC e^{-\iC \omega t} \int_{-K} ^{+K} J_0\left(\rho \sqrt{K^2-k^2}\right )\,\left(\omega -\frac{c^2}{\omega }\,k^2 \right )\,S(k)\,e^{ikz} \dd k \right ],
\ee
where $K := \omega /c$. In the case with a (discrete) frequency spectrum, one just has to sum each component: (\ref{Bphi_mono}), (\ref{Erho_mono}), or (\ref{Ez_mono}), over the different frequencies $\ \omega _j\ $, with the corresponding values $\ K_j=\omega _j/c\ $ and spectra $\ S_j=S_j(k)\ (-K_j\leq k \leq +K_j)\ $ --- as with a scalar wave (\ref{psi_avec_spectre}):
\be\label{Bphi_poly}
B_{\phi } = \sum _{j=1} ^{N_\omega } B_{\phi \,\omega_j \, S_j},
\ee
\be\label{Erho_poly}
E_{\rho } = \sum _{j=1} ^{N_\omega } E_{\rho \,\omega_j \, S_j},
\ee
\be\label{Ez_poly}
E_{z } = \sum _{j=1} ^{N_\omega } E_{z \,\omega_j \, S_j}.
\ee

\section{From Maxwell fields to scalar waves}\label{Max_to_scal}

Now an important question arises: Do the GAZR solutions generate all axisymmetric time-harmonic solutions of the Maxwell equations (in which case, by summation on frequencies, they would generate all axisymmetric solutions of the Maxwell equations)? That is: {\it let $\ ({\bf A}, {\bf E}, {\bf B})\ $ be any time-harmonic axisymmetric solution of the free Maxwell equations. Can one find a GAZR1 solution and a GAZR2 solution, whose sum give just that starting solution?} \\

\hypertarget{Complementarity}{Note from Eqs. (\ref{Bfi})--(\ref{Ez}) and (\ref{dual})} that {\it the GAZR1 solution and the GAZR2 solution are complementary:} in cylindrical coordinates, the GAZR1 solution provides non-zero components $B_\phi , E_\rho$, $E_z$, the other components $E_\phi $, $B_\rho $, $B_z$ being zero --- and the exact opposite is true for the GAZR2 solution. In view of this complementarity, we can consider separately the two sets of components: $B_\phi , E_\rho$, $E_z$ on one side, and $E_\phi $, $B_\rho $, $B_z$ on the other side. 

\subsection{Sufficient conditions for the existence of the decomposition}

For the ``GAZR1" solution, which gives non-zero values to the first among the two sets of components just mentioned, we have the following result:

\paragraph{Proposition 1.}\label{Proposition 1} {\it Let $({\bf A}, {\bf E}, {\bf B})$ be any time-harmonic axisymmetric solution of the free Maxwell equations. In order that a time-harmonic axisymmetric solution $(A_{1z},B_{1 \phi}, E_{1 \rho }, E_{1 z }; E_{1 \phi}= B_{1 \rho }= B_{1 z }=0  )$, of the form (\ref{Bfi})--(\ref{Ez}), and having the same frequency $\omega $ as the starting solution $({\bf A}, {\bf E}, {\bf B})$, be such that \Couleur{$B_{1 \phi} = B_{\phi}, \ E_{1 \rho }=E_{\rho }, \ E_{1 z }= E_{z }$}, it is sufficient that we have just} 
\be\label{B1fi=Bfi}
B_{1 \phi} = B_{\phi}.
\ee

\vspace{2mm}
\noi {\it Proof.} Let $A_{1z}(t,\rho ,z)$ be a time-harmonic axisymmetric solution of the wave equation, with frequency $\omega $, and assume that $B_{1 \phi}$ as defined by Eq. (\ref{Bfi}) [with   $A_{1z}$ in the place of  $A_{z}$] is equal to $B_{\phi}$, where ${\bf B}$ is defined by Eq. (\ref{B_from_A}). I.e., assume that
\be\label{for_B1fi=Bfi}
-\frac{\partial A_{1 z}}{\partial \rho } = \frac{\partial A_\rho }{\partial z }-\frac{\partial A_z}{\partial \rho }.
\ee
Denoting by ${\bf A}_1:=A_{1 z} {\bf e}_z$ the vector potential that provides the GAZR1 solution $(B_{1 \phi}, E_{1 \rho }, E_{1 z }; E_{1 \phi}= B_{1 \rho }= B_{1 z }=0  )$, let us compute $E_\rho  - E_{1 \rho } $ and $E_z - E_{1 z}$. We have by Eq. (\ref{E_from_A}):
\be\label{E-E_1}
\frac{\omega }{\iC c^2}\left({\bf E}-{\bf E}_1 \right ) =  \nabla \left ( \mathrm{div} {\bf A}' \right ) + \frac{\omega^2 }{c^2} {\bf A}',
\ee
where
\be\label{def_A'}
\qquad {\bf A}':={\bf A}- {\bf A}_1 :={\bf A}-A_{1 z} {\bf e}_z.
\ee
In order that the vector potential ${\bf A}$ of the {\it a priori} given solution $({\bf A}, {\bf E}, {\bf B})$ be axisymmetric, its components $A_\rho ,\, A_\phi ,\,A_z$ must depend only on $t,\rho ,z$, i.e., be independent of $\phi $. Therefore:
\be\label{div A}
\mathrm{div} {\bf A} = \frac{1}{\rho } \frac{\partial (\rho A_\rho )}{\partial \rho } + \frac{\partial A_z}{\partial z} ,
\ee
and, using this and (\ref{def_A'}):
\be
\mathrm{div} {\bf A}' = \frac{1}{\rho } \frac{\partial (\rho A_\rho )}{\partial \rho } + \frac{\partial A'_z}{\partial z}.
\ee
Hence, in (\ref{E-E_1}), we have
\bea\label{div_A'}
\nabla \left (\mathrm{div} {\bf A}' \right ) & = & \nabla \left ( \frac{\partial A_\rho }{\partial \rho } + \frac{A_\rho }{\rho }  + \frac{\partial A'_z}{\partial z} \right ) \\
\nonumber & = & \left (\frac{\partial ^2 A_\rho }{\partial \rho ^2} +\frac{1}{\rho } \frac{\partial  A_\rho }{\partial \rho } -\frac{1}{\rho^2 }A_\rho  + \frac{\partial^2 A'_z}{\partial \rho \partial z} \right ) {\bf e}_\rho + \left ( \frac{\partial ^2 A_\rho }{\partial z \partial \rho } + \frac{1}{\rho}\frac{\partial  A_\rho }{\partial z } + \frac{\partial ^2 A'_z }{\partial z^2 }\right ) {\bf e}_z.
\eea
The radial component of the vector (\ref{E-E_1}) is thus:
\be\label{E-E1_rho}
\frac{\omega }{\iC c^2}\left({\bf E}-{\bf E}_1 \right )_\rho  = \frac{\partial ^2 A_\rho }{\partial \rho ^2} +\frac{1}{\rho } \frac{\partial  A_\rho }{\partial \rho } -\frac{A_\rho}{\rho^2 }  + \frac{\partial^2 A_z}{\partial \rho \partial z} - \frac{\partial^2 A_{1 z}}{\partial \rho \partial z} + \frac{\omega^2 }{c^2} A_\rho .
\ee
However, the vector potential ${\bf A}$ obeys the Helmholtz equation (\ref{Helmholtz}), that is for the radial component (using the fact that $\frac{\partial A_\rho }{\partial \phi }= \frac{\partial A_\phi  }{\partial \phi }\equiv 0$):
\bea\label{Helm_rho}
\left (\Delta {\bf A} \right )_\rho + \frac{\omega^2 }{c^2} A_\rho & \equiv  &\Delta A_\rho -\frac{A_\rho}{\rho^2} -\frac{2}{\rho^2}\frac{\partial A_\phi  }{\partial \phi } + \frac{\omega^2 }{c^2} A_\rho \nonumber\\
& \equiv  & \frac{\partial ^2 A_\rho }{\partial \rho ^2} +\frac{1}{\rho } \frac{\partial  A_\rho }{\partial \rho } + \frac{\partial ^2 A_\rho  }{\partial z^2 } -\frac{A_\rho}{\rho^2 } +\frac{\omega^2 }{c^2} A_\rho =0.
\eea
Inserting (\ref{Helm_rho}) into (\ref{E-E1_rho}) gives us:
\be\label{E-E1_rho_2}
\frac{\omega }{\iC c^2}\left({\bf E}-{\bf E}_1 \right )_\rho  = -\frac{\partial ^2 A_\rho }{\partial z ^2} +\frac{\partial^2 A_z}{\partial \rho \partial z} - \frac{\partial^2 A_{1 z}}{\partial \rho \partial z}= -\frac{\partial }{\partial z} \left ( \frac{\partial A_\rho }{\partial z }-\frac{\partial A_z}{\partial \rho }+\frac{\partial A_{1 z}}{\partial \rho }  \right ).
\ee
Therefore, if Eq. (\ref{for_B1fi=Bfi}) is satisfied, then we have $E_{1 \rho }= E_\rho$.\\

Similarly, from (\ref{div_A'}), the axial component of the vector (\ref{E-E_1}) is 
\be\label{E-E1_z}
\frac{\omega }{\iC c^2}\left({\bf E}-{\bf E}_1 \right )_z  = \frac{\partial ^2 A_\rho }{\partial z \partial \rho } +\frac{1}{\rho } \frac{\partial  A_\rho }{\partial z }  + \frac{\partial^2 A_z}{\partial z^2} - \frac{\partial^2 A_{1 z}}{\partial z^2} + \frac{\omega^2 }{c^2} \left ( A_z -A_{1 z} \right ).
\ee
On the other hand, the axial component of the Helmholtz equation (\ref{Helmholtz}) is
\bea\label{Helm_z}
\left (\Delta {\bf A} \right )_z + \frac{\omega^2 }{c^2} A_z & \equiv  &\Delta A_z + \frac{\omega^2 }{c^2} A_z \nonumber\\
& \equiv  & \frac{\partial ^2 A_z }{\partial \rho ^2} +\frac{1}{\rho } \frac{\partial  A_z }{\partial \rho } + \frac{\partial ^2 A_z  }{\partial z^2 }  +\frac{\omega^2 }{c^2} A_z =0.
\eea
If Eq. (\ref{for_B1fi=Bfi}) is satisfied, we have
\be
\frac{\partial ^2 A_z }{\partial \rho ^2} = \frac{\partial ^2 A_\rho }{\partial \rho  \partial z } +  \frac{\partial^2 A_{1 z}}{\partial \rho ^2} .
\ee
In Eq. (\ref{Helm_z}), we replace $\frac{\partial ^2 A_z }{\partial \rho ^2}$ by its value given on the r.h.s. above, and we replace $\frac{\partial  A_z }{\partial \rho }$ by its value given by Eq. (\ref{for_B1fi=Bfi}). This gives: 
\be\label{Helm_z_&Bfi}
 \frac{\partial ^2 A_\rho  }{\partial \rho \partial z}+ \frac{\partial ^2 A_{1 z}  }{\partial \rho ^2 }  +\frac{1}{\rho } \frac{\partial  A_\rho  }{\partial z } + \frac{1}{\rho}\frac{\partial A_{1 z}}{\partial \rho }+ \frac{\partial ^2 A_z  }{\partial z^2 }  +\frac{\omega^2 }{c^2} A_z =0.
\ee
Using this equation in Eq. (\ref{E-E1_z}), we rewrite the latter as
\be\label{E-E1_z_2}
\frac{\omega }{\iC c^2}\left({\bf E}-{\bf E}_1 \right )_z  = -\frac{\partial ^2 A_{1 z}  }{\partial \rho ^2 } - \frac{1}{\rho}\frac{\partial A_{1 z}}{\partial \rho } - \frac{\partial^2 A_{1 z}}{\partial z^2} - \frac{\omega^2 }{c^2} \,A_{1 z}.
\ee
We recognize the r.h.s as that of Eq. (\ref{Helm_z}), though with the minus sign, and with $A_{1 z}$ in the place of $A_z$. I.e., Eq. (\ref{E-E1_z_2}) is just
\be\label{E-E1_z_3}
\frac{\omega }{\iC c^2}\left({\bf E}-{\bf E}_1 \right )_z  = -\Delta A_{1 z} - \frac{\omega^2 }{c^2} A_{1 z}.
\ee
But this is zero, since $A_{1 z}$ is by assumption a time-harmonic solution of the wave equation, with frequency $\omega $. Therefore, if Eq. (\ref{for_B1fi=Bfi}) is satisfied, then we have $E_{1 z }= E_z$, too. This completes the proof of \hyperref[Proposition 1]{Proposition 1}. \hfill $\square$ \\

It follows for the dual (``GAZR2") solution:

\paragraph{Corollary 1.}\label{Corollary 1} {\it Let $({\bf A}, {\bf E}, {\bf B})$ be any time-harmonic axisymmetric solution of the free Maxwell equations. In order that a time-harmonic solution} ($A_{2z}$, $E'_{2 \phi}$, $B'_{2 \rho }$, $B'_{2 z }$; $B'_{2 \phi}= E'_{2 \rho }= E'_{2 z }=0 $) {\it with the same frequency, deduced from Eqs. (\ref{Bfi})--(\ref{Ez}) by the duality (\ref{dual}), be such that $E'_{2 \phi} = E_{\phi}, \ B'_{2 \rho }=B_{\rho }, \ B'_{2 z }= B_{z }$, it is sufficient that we have just}
\be\label{E'2fi=Efi}
E'_{2 \phi}= E_\phi .
\ee\\

\noi {\it Proof.} The GAZR2 solution $(A_{2z},E'_{2 \phi}, B'_{2 \rho }, B'_{2 z }; B'_{2 \phi}= E'_{2 \rho }= E'_{2 z }=0  )$ is deduced from the GAZR1 solution $(A_{2 z},B_{2 \phi}, E_{2 \rho }, E_{2 z }; E_{2 \phi}= B_{2 \rho }= B_{2 z }=0  )$, associated with the same potential $A_{2 z}$, by the duality relation (\ref{dual}). Suppose that Eq. (\ref{E'2fi=Efi}) is satisfied. With the starting solution $({\bf A}, {\bf E}, {\bf B})$ of the free Maxwell equations, we may associate another solution, by the inverse duality:\\
\be\label{inverse_dual}
{\bf \widetilde{ B}}=\frac{1}{c}{\bf E}, \quad {\bf \widetilde{ E}}=-c {\bf B}.
\ee
The assumed relation (\ref{E'2fi=Efi}) means that
\be\label{B2fi=Btildefi}
B_{2 \phi}= \widetilde{ B}_\phi.
\ee
Indeed, by applying successively (\ref{dual})$_1$, (\ref{E'2fi=Efi}), and (\ref{inverse_dual})$_1$, we obtain:
\be
B_{2 \phi} = \frac{1}{c} E'_{2 \phi} = \frac{1}{c} E_\phi = \widetilde{ B}_\phi.
\ee
In turn, the relation (\ref{B2fi=Btildefi}) means that we may apply \hyperref[Proposition 1]{Proposition 1} to the GAZR1 solution  $(A_{2 z},B_{2 \phi},...)$ and the solution $({\bf \widetilde{ A}}, {\bf \widetilde{ E}}, \widetilde{ {\bf B}})$. 
\footnote{\label{Atilde}\
The explicit expression of the corresponding vector potential ${\bf \widetilde{ A}}$ as function of $({\bf A}, {\bf E}, {\bf B})$ is not needed: only the existence of an axisymmetric ${\bf \widetilde{ A}}$, such that ${\bf \widetilde{ B}} = \mathrm{rot} {\bf \widetilde{ A}}$, is needed. Precisely, ${\bf \widetilde{ A}}$ is got as a solution of the PDE \ $\mathrm{rot} {\bf \widetilde{ A}} = {\bf \widetilde{ B}}$. (Such a solution always exists in a topologically trivial domain, thus in particular if the domain is the whole space.) Hence ${\bf \widetilde{ A}}$ can indeed be chosen axisymmetric, i.e. with its components in cylindrical coordinates being independent of $\phi $, because with this choice the independent variables of the PDE \ $\mathrm{rot} {\bf \widetilde{ A}} = {\bf \widetilde{ B}}$ \ simply do not include $\phi $ --- since ${\bf \widetilde{ B}}=\frac{1}{c}{\bf E}$ is axisymmetric, as is $ {\bf E}$  by assumption.
}
Thus, \hyperref[Proposition 1]{Proposition 1} tells us here that, since $B_{2 \phi}= \widetilde{ B}_\phi$, we have also 
\be
E_{2 \rho }= \widetilde{ E}_\rho  \quad \mathrm{and}\quad E_{2 z}= \widetilde{ E}_z,
\ee
i.e., in view of (\ref{dual})$_2$ and (\ref{inverse_dual})$_2$:
\be
-cB'_{2 \rho }= -cB_\rho  \quad \mathrm{and}\quad -cB'_{2 z}= -cB_z.
\ee
This proves \hyperref[Corollary 1]{Corollary 1}. \hfill $\square$ \\

\subsection{Generality of the decomposition}

\paragraph{Proposition 2.}\label{Proposition 2} {\it Let $({\bf A}, {\bf E}, {\bf B})$ be any time-harmonic axisymmetric solution of the free Maxwell equations. There exists a time-harmonic axisymmetric solution $A_{2 z}$ of the wave equation, with the same frequency, such that the associated GAZR2 solution $({\bf E}'_2, {\bf B}'_2)$, deduced from $A_{2 z}$ by Eqs. (\ref{Bfi})--(\ref{Ez}) followed by the duality transformation (\ref{dual}), satisfy} 
\be\label{GAZR2_OK}
E'_{2 \phi} = E_{\phi}, \quad B'_{2 \rho }=B_{\rho }, \quad B'_{2 z }= B_{z }.
\ee

\vspace{2mm}
\noi {\it Proof.} In view of \hyperref[Corollary 1]{Corollary 1}, we merely have to prove that there exists a time-harmonic axisymmetric solution $A_{2 z}$ of the wave equation, such that Eq. (\ref{E'2fi=Efi}) is satisfied. From Eqs. (\ref{E_from_A}) and (\ref{div A}), we have 
\be\label{Efi}
E_\phi = \iC \omega  A_\phi .
\ee
Using this with Eqs. (\ref{Bfi}) and (\ref{dual})$_1$, we may rewrite the sought-for relation (\ref{E'2fi=Efi}) as: 
\be\label{Efi-2}
-\frac{\partial A_{2 z}}{\partial \rho } = \frac{\iC \omega}{c}  A_\phi. 
\ee 
This equation can be solved by a quadrature:
\be\label{A_2z}
 A_{2 z}(t,\rho ,z) = h(t,z) + \int_{\rho _0} ^\rho -\frac{\iC \omega}{c} A_\phi (t,\rho ',z)\,\dd \rho '.
\ee
We thus have to find out if it is possible to determine the function $h$ so that $A_{2 z}$ given by (\ref{A_2z}) obey the wave equation. Moreover, the unknown function $A_{2 z}$ must have a harmonic time dependence with frequency $\omega $ as has $A_\phi $, i.e., 
\be\label{psi}
A_{2 z}(t,\rho ,z) = \psi (\rho ,z) \,e^{-\iC \omega t},\quad A_\phi (t,\rho ,z) = \hat{A}_\phi (\rho ,z) \,e^{-\iC \omega t},
\ee
so we must have $h(t,z)=e^{-\iC \omega t} g(z)$, too. Hence we may rewrite (\ref{A_2z}) as
\be\label{psi=}
 \psi (\rho ,z) = g(z) + \int_{\rho _0} ^\rho -\frac{\iC \omega}{c} \hat{A}_\phi (\rho ',z)\,\dd \rho ',
\ee
and now the question is to know if $g$ can be determined so that $\psi $ obey the scalar Helmholtz equation, i.e. [cf. Eq. (\ref{Helm_z})]:
\be\label{Helm_scal}
\mathcal{H} \psi := \Delta \psi +\frac{\omega^2 }{c^2} \psi \equiv \frac{\partial ^2 \psi }{\partial \rho ^2} +\frac{1}{\rho } \frac{\partial  \psi }{\partial \rho } + \frac{\partial ^2 \psi  }{\partial z^2 }  +\frac{\omega^2 }{c^2} \psi =0
\ee
--- knowing that $A_\phi$ or $\hat{A}_\phi$ does obey the $\phi $ component of the vector Helmholtz equation (\ref{Helmholtz}), i.e.:
\bea\label{Helm_phi}
\left (\Delta {\bf \hat{A}} \right )_\phi  + \frac{\omega^2 }{c^2} \hat{A}_\phi  & \equiv  &\Delta \hat{A}_\phi  -\frac{\hat{A}_\phi }{\rho^2} +\frac{2}{\rho^2}\frac{\partial \hat{A}_\rho  }{\partial \phi } + \frac{\omega^2 }{c^2} \hat{A}_\phi  \nonumber\\
& \equiv  &       
 \frac{\partial ^2 \hat{A}_\phi }{\partial \rho ^2} +\frac{1}{\rho } \frac{\partial  \hat{A}_\phi }{\partial \rho } + \frac{\partial ^2 \hat{A}_\phi  }{\partial z^2 } -\frac{\hat{A}_\phi}{\rho^2 } +\frac{\omega^2 }{c^2} \hat{A}_\phi = 0.
\eea
We have from Eqs. (\ref{Efi-2}) and (\ref{psi}):
\be\label{dpsi_drho}
\frac{\partial \psi }{\partial \rho } = -\frac{\iC \omega}{c}  \hat{A}_\phi, 
\ee 
hence
\be\label{d2psi_drho2}
\frac{\partial^2 \psi }{\partial \rho^2 } = -\frac{\iC \omega}{c}  \frac{\partial \hat{A}_\phi }{\partial \rho } . 
\ee 
And we get from (\ref{psi=}):
\be\label{dpsi_dz}
 \frac{\partial \psi}{\partial z} = \frac{\dd g}{\dd z} - \int_{\rho _0} ^\rho \frac{\iC \omega}{c} \frac{\partial \hat{A}_\phi}{\partial z} (\rho ',z)\,\dd \rho ',
\ee
whence
\be\label{d2psi_dz2}
 \frac{\partial^2 \psi}{\partial z^2} = \frac{\dd ^2 g}{\dd z^2} - \int_{\rho _0} ^\rho \frac{\iC \omega}{c} \frac{\partial^2 \hat{A}_\phi}{\partial z^2} (\rho ',z)\,\dd \rho '.
\ee
Entering Eqs. (\ref{dpsi_drho}), (\ref{d2psi_drho2}) and (\ref{d2psi_dz2})  into (\ref{Helm_scal})$_1$, we obtain:
\bea
\mathcal{H} \psi & = & -\frac{\iC \omega}{c}  \frac{\partial \hat{A}_\phi }{\partial \rho } -\frac{\iC \omega}{c}  \frac{\hat{A}_\phi }{\rho } + \frac{\dd ^2 g}{\dd z^2} - \int_{\rho _0} ^\rho \frac{\iC \omega}{c} \frac{\partial^2 \hat{A}_\phi}{\partial z^2} (\rho ',z)\,\dd \rho ' \\ \nonumber & & + \frac{\omega^2 }{c^2}  \left (g -\frac{\iC \omega}{c} \int_{\rho _0} ^\rho \hat{A}_\phi (\rho ',z)\,\dd \rho '\right ).
\eea
Therefore, the scalar Helmholtz equation (\ref{Helm_scal})$_2$ rewrites as
\be
\frac{\dd ^2 g}{\dd z^2} + \frac{\omega^2 }{c^2} g = \frac{\iC \omega}{c} \left [ \frac{\partial \hat{A}_\phi }{\partial \rho } + \frac{\hat{A}_\phi }{\rho } + \int_{\rho _0} ^\rho \left(\frac{\partial^2 \hat{A}_\phi}{\partial z^2} + \frac{\omega^2 }{c^2} \hat{A}_\phi \right ) \dd \rho ' \right ],
\ee
or, using (\ref{Helm_phi}):
\be\label{Helm_scal_2}
\frac{\dd ^2 g}{\dd z^2} + \frac{\omega^2 }{c^2} g = \frac{\iC \omega}{c} \left [ \frac{\partial \hat{A}_\phi }{\partial \rho } + \frac{\hat{A}_\phi }{\rho } + \int_{\rho _0} ^\rho \left(-\frac{\partial^2 \hat{A}_\phi}{\partial \rho '^2} - \frac{1 }{\rho '} \frac{\partial \hat{A}_\phi}{\partial \rho '} + \frac{\hat{A}_\phi}{ \rho '^2} \right ) \dd \rho ' \right ].
\ee
An integration by parts gives us:
\be
\int_{\rho _0} ^\rho \left(-\frac{\partial^2 \hat{A}_\phi}{\partial \rho '^2}  + \frac{\hat{A}_\phi}{ \rho '^2} \right ) \dd \rho ' = - \left [ \frac{\partial \hat{A}_\phi}{\partial \rho '} +\frac{\hat{A}_\phi}{ \rho '}  \right ]_{\rho _0} ^\rho  + \int_{\rho _0} ^\rho  \frac{1 }{\rho '} \frac{\partial \hat{A}_\phi}{\partial \rho '} \dd \rho ',
\ee
so Eq. (\ref{Helm_scal_2}) rewrites as
\be\label{Helm_scal_3}
\frac{\dd ^2 g}{\dd z^2} + \frac{\omega^2 }{c^2} g = \frac{\iC \omega}{c} \left [  \frac{\partial \hat{A}_\phi}{\partial \rho } (\rho _0,z) + \frac{\hat{A}_\phi (\rho _0,z)}{\rho_0 }\right ].
\ee
As is well known and easy to check, this very ordinary differential equation can be solved explicitly by the method of variation of constants. (The general solution $g$ of (\ref{Helm_scal_3}) depends linearly on two arbitrary constants.) And by construction, any among the solutions $g$ of (\ref{Helm_scal_3}) is such that, with that $g$, the function $\psi $ in Eq. (\ref{psi=}) obeys the scalar Helmholtz equation (\ref{Helm_scal}). This proves \hyperref[Proposition 2]{Proposition 2}. \hfill $\square$ \\

\paragraph{Corollary 2.}\label{Corollary 2} {\it Let $({\bf A}, {\bf E}, {\bf B})$ be any time-harmonic axisymmetric solution of the free Maxwell equations. There exists a time-harmonic axisymmetric solution $A_{1 z}$ of the wave equation, with the same frequency, such that the associated GAZR1 solution $({\bf E}_1, {\bf B}_1)$, deduced from $A_{1 z}$ by Eqs. (\ref{Bfi})--(\ref{Ez}), satisfy} 
\be\label{GAZR1_OK}
B_{1 \phi} = B_{\phi}, \quad E_{1 \rho } = E_{\rho }, \quad E_{1 z }= E_{z }.
\ee

\vspace{2mm}
\noi {\it Proof.} Let $A_{1 z}$ be a time-harmonic axisymmetric solution of the wave equation, and consider:\\

(i) the GAZR1 solution defined from $A_{1 z}$ by Eqs. (\ref{Bfi})--(\ref{Ez}) (thus with $A_{1 z},\ B_{1 \phi }, \ E_{1 \rho}, \ E_{1 z}, ... $ instead of $A_{z},\ B_{\phi }, \ E_{\rho}, \ E_{z}, ...$ respectively).\\

(ii) the GAZR2 solution defined from the same $A_{1 z}$ by applying the duality (\ref{dual}) to the said GAZR1 solution:
\be
E'_{1 \phi } := cB_{1 \phi }, \quad B'_{1 \rho } := -\frac{1}{c} E_{1 \rho },\quad B'_{1 z } := -\frac{1}{c} E_{1 z }, 
\ee
\be
E'_{1 \rho  } := cB_{1 \rho  }=0, \quad E'_{1 z } := c B_{1 z }=0,\quad B'_{1 \phi  } := -\frac{1}{c} E_{1 \phi  }=0.
\ee
 On the other hand, consider the free Maxwell field \ $( {\bf E}', {\bf B}')$ deduced from the given time-harmonic axisymmetric solution \ $( {\bf E}, {\bf B})$ \ of the free Maxwell equations by the same duality relation:
 \be\label{dual-2}
{\bf E}' := c{\bf B}, \quad {\bf B}' := -{\bf E}/c.
\ee
Just in the same way as it was shown in Note \ref{Atilde}, we know that a vector potential ${\bf A}'$ such that ${\bf B}' =  \mathrm{rot} {\bf A}'$ does exist and can be chosen to be axisymmetric (and is indeed chosen so) --- as are ${\bf E}$ and ${\bf B}$, and hence ${\bf E}'$ and ${\bf B}'$. Clearly, the sought-for relation (\ref{GAZR1_OK}) is equivalent to
\be\label{GAZR2_1_OK}
E'_{1 \phi} = E'_{\phi}, \quad B'_{1 \rho }=B'_{\rho }, \quad B'_{1 z }= B'_{z }.
\ee
Therefore, the existence of $A_{1 z}$ as in the statement of \hyperref[Corollary 2]{Corollary 2} is ensured by \hyperref[Proposition 2]{Proposition 2}. \hfill $\square$ \\

Accounting for \hyperref[Proposition 2]{Proposition 2} and for \hyperref[Corollary 2]{Corollary 2}, and remembering the \hyperlink{Complementarity}{``complementarity"} of the GAZR1 and GAZR2 solutions, we thus can answer positively to the question asked at the beginning of \hyperref[Max_to_scal]{this section}:\\

\paragraph{Theorem.}\label{Theorem} {\it Let $({\bf A}, {\bf E}, {\bf B})$ be any time-harmonic axisymmetric solution of the free Maxwell equations. There exist a unique \hyperlink{GAZR}{GAZR1} solution $({\bf E}_1, {\bf B}_1)$ and a unique \hyperlink{GAZR}{GAZR2} solution $({\bf E}'_2, {\bf B}'_2)$, both with the same frequency as has $({\bf A}, {\bf E}, {\bf B})$, and whose sum gives just that solution:}
\be
{\bf E} = {\bf E}_1 + {\bf E}'_2,\qquad {\bf B} = {\bf B}_1 + {\bf B}'_2.
\ee

\noi {\it Remark.} Thus, the uniqueness of the representation concerns the electric and magnetic fields. It of course does not concern the potentials ${\bf A}_1 = A_{1 z} {\bf e}_z$ and ${\bf A}_2 = A_{2 z}{\bf e}_z$ that generate respectively $({\bf E}_1, {\bf B}_1)$ and $({\bf E}'_2, {\bf B}'_2)$.

  \section{Discussion and conclusion}\label{Conclusion}

The authors of Ref. \cite{GAZR2014} introduced two classes of axisymmetric solutions of the free Maxwell equations, and they showed that these two classes allow one to obtain in explicit form nonparaxial EM beams. It has been proved here that, by combining these two classes, one can define a method that allows one to get {\it all} totally propagating, time-harmonic, axisymmetric free Maxwell fields --- and thus, by the appropriate summation on frequencies, all totally propagating axisymmetric free Maxwell fields. This method results immediately from the \hyperref[Theorem]{Theorem} just above, and from the general form (\ref{psi_monochrom}) of  a totally propagating, time-harmonic, axisymmetric solution of the scalar wave equation. However, that theorem is not an obviously expected result, and its proof is not immediate. We thus have now a constructive method to obtain all totally propagating axisymmetric free Maxwell fields. Namely, considering a discrete frequency spectrum $(\omega _j)_{j=1,...,N_\omega }$ for simplicity: there are $2\,N_\omega $ functions, $k \mapsto S_j(k)$, and $k \mapsto S'_j(k)\ \ (j=1,...,N_\omega )$, such that the components $B_\phi , E_\rho , E_z$ of the field are given by Eqs. (\ref{Bphi_poly}), (\ref{Erho_poly}), (\ref{Ez_poly}) respectively --- while the components $E_\phi , B_\rho , B_z$ are given by these same equations applied with the primed spectra $S'_j$, followed by the duality transformation (\ref{dual}).\\

In a forthcoming work, we shall apply this to model the interstellar radiation field in a disc galaxy as an (axisymmetric) exact solution of the free Maxwell equations. In this application, it is very important that, due to the present work, one knows that any (totally propagating) axisymmetric free Maxwell field can be got in this way.


\end{document}